\documentclass[aps,prc,twocolumn,floatfix,nofootinbib,showpacs,superscriptaddress]{revtex4-1}

\usepackage{amsmath,amsfonts,amssymb,bm}
\usepackage{graphicx}
\usepackage{color}
\definecolor{purple}{rgb}{0.5,0,0.5}
\definecolor{blue}{rgb}{0.0,0,0.9}

\begin{document}

\title{Interaction model for the gap equation}

\author{Si-xue Qin}
\affiliation{Department of Physics, Center for High Energy Physics and State Key Laboratory of Nuclear Physics and Technology, Peking University, Beijing 100871, China}
\affiliation{Physics Division, Argonne National Laboratory, Argonne, Illinois 60439, USA}

\author{Lei Chang}
\affiliation{Physics Division, Argonne National Laboratory, Argonne,
Illinois 60439, USA}

\author{Yu-xin Liu}
\affiliation{Department of Physics, Center for High Energy Physics and State Key Laboratory of Nuclear Physics and Technology, Peking University, Beijing 100871, China}
\affiliation{Center of Theoretical Nuclear Physics, National
Laboratory of Heavy Ion Accelerator, Lanzhou 730000, China}

\author{Craig D.~Roberts}
\affiliation{Department of Physics, Center for High Energy Physics and State Key Laboratory of Nuclear Physics and Technology, Peking University, Beijing 100871, China}
\affiliation{Physics Division, Argonne National Laboratory, Argonne, Illinois 60439, USA}
\affiliation{Institut f\"ur Kernphysik, Forschungszentrum J\"ulich, D-52425 J\"ulich, Germany}
\affiliation{Department of Physics, Illinois Institute of Technology, Chicago, Illinois 60616-3793, USA}

\author{David J.~Wilson}
\affiliation{Physics Division, Argonne National Laboratory, Argonne, Illinois 60439, USA}


\begin{abstract}
We explain a form for the rainbow-ladder kernel whose momentum-dependence is consonant with modern DSE- and lattice-QCD results, and assess its capability as a tool in hadron physics.  In every respect tested, this form produces results for observables that are at least equal to the best otherwise obtained in a comparable approach.  Moreover, it enables the natural extraction of a monotonic running-coupling and -gluon-mass.  
\end{abstract}


\maketitle

%
Solving quantum chromodynamics (QCD) presents a fundamental problem: never before have we been confronted by a theory whose elementary excitations are not those degrees-of-freedom readily accessible via experiment; i.e., are \emph{confined}.  Moreover, there are reasons to believe that QCD generates forces which are so strong that $<2$\% of a nucleon's mass can be attributed to the current-quark masses that appear in QCD's Lagrangian; viz., forces capable of generating mass from nothing, a phenomenon known as dynamical chiral symmetry breaking (DCSB).  Neither confinement nor DCSB is apparent in QCD's Lagrangian.  Yet they are dominant in determining the observable characteristics of real-world QCD.  The physics of hadrons is ruled by emergent phenomena, such as these, which can only be explained through the methods of nonperturbative quantum field theory.

Confinement and DCSB are long-distance phenomena.  Their understanding requires elucidation of the infrared behaviour of the interaction between quarks and gluons.  Much is misapprehended about confinement.   It is therefore important to state that the static potential measured in simulations of lattice-QCD is not related in any known way to the question of light-quark confinement.  Light-quark creation and annihilation effects are fundamentally nonperturbative in QCD.  Hence it is impossible in principle to compute a potential between two light quarks.  On the other hand, confinement can be related to the analytic properties of QCD's Schwinger functions \cite{Krein:1990sf,Roberts:1994dr,Roberts:2007ji}, so the question of light-quark confinement may be translated into the challenge of charting the infrared behavior of QCD's $\beta$-function.

The gap equation is a starting point for analyses of DCSB and confinement \cite{Roberts:1994dr}; and it is the basis for formulating a symmetry-preserving Poincar\'e-covariant treatment of bound states through Bethe-Salpeter and Faddeev equations \cite{Maris:2003vk,Roberts:2007jh}.  The gap equation can be written
\begin{eqnarray}
\lefteqn{
\nonumber S(p)^{-1} = Z_2 \,(i\gamma\cdot p + m^{\rm bm})}\\
&& + Z_1 \int^\Lambda_q\!\! g^2 D_{\mu\nu}(p-q)\frac{\lambda^a}{2}\gamma_\mu S(q) \frac{\lambda^a}{2}\Gamma_\nu(q,p) ,
\label{gendseN}
\end{eqnarray}
where: $D_{\mu\nu}$ is the gluon propagator; $\Gamma_\nu$, the quark-gluon vertex; $\int^\Lambda_q$, a symbol that represents a Poincar\'e invariant regularization of the four-dimensional integral, with $\Lambda$ the regularization mass-scale; $m^{\rm bm}(\Lambda)$, the current-quark bare mass; and $Z_{1,2}(\zeta^2,\Lambda^2)$, respectively, the vertex and quark wave function renormalisation constants, with $\zeta$ the renormalisation point. 
The gap equation's solution is the dressed-quark propagator,
\begin{equation}
S(p) 
=\frac{Z(p^2,\zeta^2)}{i\gamma\cdot p + M(p^2)}\,,
\label{SgeneralN}
\end{equation}
wherein the mass function, $M(p^2)$, is independent of the renormalisation point.  (We use a Euclidean metric \cite{Chang:2011}.)

In realistic studies the model input is expressed in a statement about the nature of the gap equation's kernel at infrared momenta, since the behaviour at momenta $k^2\gtrsim 2\,$GeV$^2$ is fixed by perturbation theory and the renormalisation group \cite{Jain:1993qh,Maris:1997tm}.  In rainbow-ladder (RL) truncation, which is leading-order in the most widely used scheme \cite{Munczek:1994zz,Bender:1996bb}, this amounts to writing ($k=p-q$)
\begin{eqnarray}
\nonumber
\lefteqn{
Z_1 g^2 D_{\mu\nu}(k) \Gamma_\nu(q,p) = k^2 {\cal G}(k^2)
D^{\rm free}_{\mu\nu}(k) \gamma_\nu}\\
&=&   \left[ k^2 {\cal G}_{\rm IR}(k^2) + 4\pi \tilde\alpha_{\rm pQCD}(k^2) \right]
D^{\rm free}_{\mu\nu}(k) \gamma_\nu ,
\label{rainbowdse}
\end{eqnarray}
wherein $D^{\rm free}_{\mu\nu}(k)$ is the Landau-gauge free-gauge-boson propagator; 
$\tilde\alpha_{\rm pQCD}(k^2)$ is a bounded, monotonically-decreasing regular continuation of the perturbative-QCD running coupling to all values of spacelike-$k^2$; and ${\cal G}_{\rm IR}(k^2)$ is an \emph{Ansatz} for the interaction at infrared momenta: ${\cal G}_{\rm IR}(k^2)\ll \tilde\alpha_{\rm pQCD}(k^2)$ $\forall k^2\gtrsim 2\,$GeV$^2$.  The form of ${\cal G}_{\rm IR}(k^2)$ determines whether confinement and/or DCSB are realised in solutions of the gap equation.  (Landau gauge is used for many reasons \protect\cite{Bashir:2009fv}, for example, it is: a fixed point of the renormalisation group; that gauge for which sensitivity to model-dependent differences between \emph{Ans\"atze} for the fermion--gauge-boson vertex are least noticeable; and a covariant gauge, which is readily implemented in numerical simulations of lattice regularised QCD \protect\cite{Cucchieri:2011aa}.)

The capacity of the Dyson-Schwinger equations (DSEs) to unify the explanation of a wide range of meson and baryon observables entails that the pointwise behaviour of ${\cal G}_{\rm IR}(k^2)$ can be constrained through feedback between experiment and theory.  This is a practical means by which to develop insight into the momentum-dependence of QCD's $\beta$-function \cite{Roberts:2007ji,Aznauryan:2009da,Chang:2011}.

Following work on confinement \cite{Munczek:1983dx,Burden:1991gd}, the interaction at small-$k^2$ has often been expressed as either an integrable infrared singularity, typically ${\cal G}_{\rm IR}(k^2) \propto \delta^4(k)$, or a finite-width approximation to it \cite{Jain:1993qh,Maris:1997tm,Frank:1995uk,Burden:1996vt,Maris:1999nt}.  The following approximation was used in Ref.\,\cite{Jain:1993qh}:
\begin{equation}
\label{deltaJM}
\delta^4(k) \stackrel{\omega \sim 0}{\approx} \frac{1}{\pi^2}\frac{1}{\omega^4} {\rm e}^{-k^2/\omega^2},
\end{equation}
whereas the $\delta^4$-function itself was used in Ref.\,\cite{Frank:1995uk}.  Neither study directly sampled the solutions of the DSEs at complex values of their arguments.  This changed with Ref.\,\cite{Maris:1997tm}, which therefore required better control of numerical procedures and hence employed an equally weighted combination of the $\delta^4$-function and the following finite width representation:
\begin{equation}
\label{deltaMR}
\delta^4(k) \stackrel{\omega \sim 0}{\approx} \frac{1}{2\pi^2}\frac{1}{\omega^6} \, k^2 \, {\rm e}^{-k^2/\omega^2}.
\end{equation}
The material difference between this form and Eq.\,(\ref{deltaJM}) is the inclusion of a multiplicative factor of $k^2$.  It was introduced solely in order to tame singularities encountered in the numerical treatment of the transverse projection operator $k^2 D^{\rm free}_{\mu\nu}(k)$.

Desiring additional simplifications in numerical analysis, the $\delta^4$-function component of ${\cal G}_{\rm IR}(k^2)$ was completely eliminated in Ref.\,\cite{Maris:1999nt}, leaving the infrared behaviour to be described by Eq.\,(\ref{deltaMR}) alone; viz., with $s=k^2$,
\begin{equation}
\label{CalGMT}
{\cal G}(s) = \frac{4 \pi^2}{\omega^6} D s \, {\rm e}^{-s/\omega^2}
+ \frac{8 \pi^2 \gamma_m}{\ln [ \tau + (1+s/\Lambda_{\rm QCD}^2)^2]} {\cal F}(s),
\end{equation}
where: $\gamma_m = 12/(33-2 N_f)$, $N_f=4$, $\Lambda_{\rm QCD}=0.234\,$GeV; $\tau={\rm e}^2-1$; and ${\cal F}(s) = \{1 - \exp(-s/[4 m_t^2])\}/s$, $m_t=0.5\,$GeV.  This form of the interaction preserves the one-loop renormalisation-group behavior of QCD in the gap equation; and has since been employed extensively in the successful prediction and explanation of hadron observables; e.g., Refs.\,\cite{Maris:2003vk,Holl:2005vu,Bhagwat:2006xi,Bhagwat:2006pu,Eichmann:2008ae,%
Eichmann:2008ef,Goecke:2010if,Nguyen:2011jy,Eichmann:2011vu,Mader:2011zf,Blank:2011qk}.

There is an aspect of the interaction in Eq.\,(\ref{CalGMT}) that is usually ignored; namely, it produces a kernel for the gap equation which possesses a zero at a small timelike value of $k^2$, and rapidly becomes very large and negative as the magnitude of the timelike momentum is increased.  For example, with typical values of the model parameters ($D\omega = (0.72\,{\rm GeV})^3$, $\omega=0.4\,$GeV):\footnote{Predictions for numerous pseudoscalar- and vector-meson observables are approximately independent of $\omega$ on the domain $\omega \in [0.3, 0.5]$ so long as one maintains $D\omega =\,$constant \protect\cite{Maris:2002mt}.}
\begin{equation}
{\cal G}(s) \stackrel{\mbox{Eq.\,(\protect\ref{CalGMT})}}{=} 0 \; {\rm for} \; s=-(0.046\,{\rm GeV})^2;
\end{equation}
the magnitude of ${\cal G}(s<0)$ exceeds its largest spacelike value at $s=-(0.22\,{\rm GeV})^2$; and $|{\cal G}(s<0)<0|$ grows faster than exponentially with decreasing $s$.

These facets of the behaviour produced by Eq.\,(\ref{CalGMT}) are in stark conflict with the results of modern DSE and lattice studies; viz., the gluon propagator is a bounded, regular function of spacelike momenta, which achieves its maximum value on this domain at $k^2=0$ \cite{Bowman:2004jm,Aguilar:2009nf,Aguilar:2010gm}, and the dressed-quark-gluon vertex does not possess any structure which can qualitatively alter this behaviour \cite{Skullerud:2003qu,Bhagwat:2004kj}.  It is thus long overdue to reconsider a functional form whose sole \emph{raison d'\^{e}tre} was numerical expediency.

We therefore choose to explore the capacity of
\begin{equation}
\label{CalGQC}
{\cal G}(s) = \frac{8 \pi^2}{\omega^4} D \, {\rm e}^{-s/\omega^2}
+ \frac{8 \pi^2 \gamma_m}{\ln [ \tau + (1+s/\Lambda_{\rm QCD}^2)^2]} {\cal F}(s),
\end{equation}
as a tool to compute and connect hadron observables.  This is readily done now owing to improved numerical methods for coping with DSE solutions at complex values of their arguments \cite{Krassnigg:2009gd}.  NB.\ Equation (\ref{CalGQC}) cannot be expressed via a non-negative spectral density \cite{Roberts:2007ji}.

Regarding renormalisation of the gap equation, we follow precisely the procedures of Refs.\,\cite{Maris:1997tm,Maris:1999nt} and use the same renormalisation point; i.e., $\zeta=19\,$GeV.  With gap equation solutions in hand for various quark flavours, one can solve homogeneous Bethe-Salpeter equations (BSEs) for meson Bethe-Salpeter amplitudes and therefrom compute observable meson properties.  For example, in the isospin-symmetric limit the pion Bethe-Salpeter amplitude is obtained via\footnote{We actually include a factor of $1/Z_2^2$ on the left-hand-side of both Eqs.\,(\protect\ref{CalGMT}) and (\protect\ref{CalGQC}), which additional improvement ensures multiplicative renormalisability in solutions of the gap and Bethe-Salpeter equations \protect\cite{Bloch:2002eq}.}
\begin{eqnarray}
\nonumber
\Gamma_\pi(k;P) &=&  - \int_q^\Lambda
{\cal G}((k-q)^2)\, (k-q)^2 \, D_{\mu\nu}^{\rm free}(k-q)\\
&& \times
\frac{\lambda^a}{2}\gamma_\mu S(q_+)\Gamma_\pi(q;P) S(q_-) \frac{\lambda^a}{2}\gamma_\nu ,
\end{eqnarray}
where $S(\ell)$ is the $u=d$-quark propagator, $P^2=-m_\pi^2$, $k$ is the relative momentum between the constituents, and one can choose $q_\pm = q\pm P/2$ without loss of generality in a Poincar\'e covariant approach.  This form of the BSE is symmetry-consistent with the gap equation obtained through Eq.\,(\ref{rainbowdse}) \cite{Bender:1996bb,Chang:2009zb}.  All Bethe-Salpeter amplitudes are normalised canonically (see, e.g., Eq.\,(27) in Ref.\,\cite{Maris:1997tm}).

\begin{table}[t]
\begin{center}
\begin{tabular*}
{\hsize}
{|l@{\extracolsep{0ptplus1fil}}
||l@{\extracolsep{0ptplus1fil}}
||l@{\extracolsep{0ptplus1fil}}
|l@{\extracolsep{0ptplus1fil}}
|l@{\extracolsep{0ptplus1fil}}
|l@{\extracolsep{0ptplus1fil}}|}\hline
Interaction & Eq.\,(\protect\ref{CalGMT}) & Eq.\,(\protect\ref{CalGQC}) & Eq.\,(\protect\ref{CalGQC}) & Eq.\,(\protect\ref{CalGQC}) & Eq.\,(\protect\ref{CalGQC}) \\\hline
$(D\omega)^{1/3}$ & 0.72 & 0.8 & 0.8 & 0.8 & 0.8 \\
$\omega$ & 0.4 & 0.4 & 0.5 & 0.6 & 0.7 \\
$m_{u,d}^\zeta$ & 0.0037 & 0.0034 & 0.0034 & 0.0034 & 0.0034 \\
$m_{s}^\zeta$ & 0.084 & 0.082 & 0.082 & 0.082 & 0.082 \\\hline
$A(0)$ & 1.58 & 2.07 & 1.70 & 1.38 & 1.16 \\
$M(0)$ & 0.50 & 0.62 & 0.52 & 0.42 & 0.29 \\
$m_\pi$ & 0.138 & 0.139 & 0.134 & 0.136 & 0.139 \\
$f_\pi$ & 0.093 & 0.094 & 0.093 & 0.090 & 0.081\\
$\rho_\pi^{1/2}$ & 0.48 & 0.49 & 0.49 & 0.49 & 0.48\\
$m_K$ & 0.496 & 0.496 & 0.495 & 0.497 & 0.503 \\
$f_K$ & 0.11 & 0.11 & 0.11 & 0.11 & 0.10 \\
$\rho_K^{1/2}$ & 0.54 & 0.55 & 0.55 & 0.55 & 0.55\\
$m_\rho$ & 0.74 & 0.76 & 0.74 & 0.72 & 0.67 \\
$f_\rho$ & 0.15 & 0.14 & 0.15 & 0.14 & 0.12 \\
$m_\phi$ & 1.07 & 1.09 & 1.08 & 1.07 & 1.05 \\
$f_\phi$ & 0.18 & 0.19 & 0.19 & 0.19 & 0.18 \\
$m_\sigma$ & 0.67 & 0.67 & 0.65 & 0.59 & 0.46 \\
$\rho_\sigma^{1/2}$ & 0.52 & 0.53 & 0.53 & 0.51 & 0.48\\\hline
\end{tabular*}
\end{center}
\caption{\label{tableresults}
Results obtained using Eq.\,(\protect\ref{CalGQC}), compared with one representative set computed using Eq.\,(\protect\ref{CalGMT}).  The current-quark masses at the renormalisation point $\zeta=0.19\,$GeV were fixed by requiring a good description of $m_{\pi,K}$.  Dimensioned quantities are reported in GeV.  NB.\ The ``$\sigma$'' listed here is not directly comparable with the lightest scalar in the hadron spectrum because the rainbow-ladder truncation is \emph{a priori} known to be a poor approximation in this channel \protect\cite{Chang:2009zb,Chang:2011ei}.}
\end{table}

In Table~\ref{tableresults} we list computed results for ground-state pseudoscalar- and vector-mesons.  The meson masses are obtained in solving the BSEs.  Valid formulae for the other quantities, all of which depend linearly on the meson Bethe-Salpeter amplitudes, are presented in Refs.\,\cite{Maris:1997tm,Ivanov:1998ms,Maris:2000ig}.  (NB.\, The products $f_\pi \rho_\pi$ and $f_K \rho_K$ describe in-pion and in-kaon condensates \cite{Maris:1997tm,Brodsky:2010xf}.)  The results show that observable properties of vector- and flavoured-pseudoscalar-mesons computed with Eq.\,(\ref{CalGQC}) are practically insensitive to variations of $\omega \in [0.4,0.6]\,$GeV so long as $D\omega=\,$constant.  Furthermore, that there is no reason to prefer Eq.\,(\ref{CalGMT}) over Eq.\,(\ref{CalGQC}).

\begin{figure}[t]
\includegraphics[clip,width=0.45\textwidth]{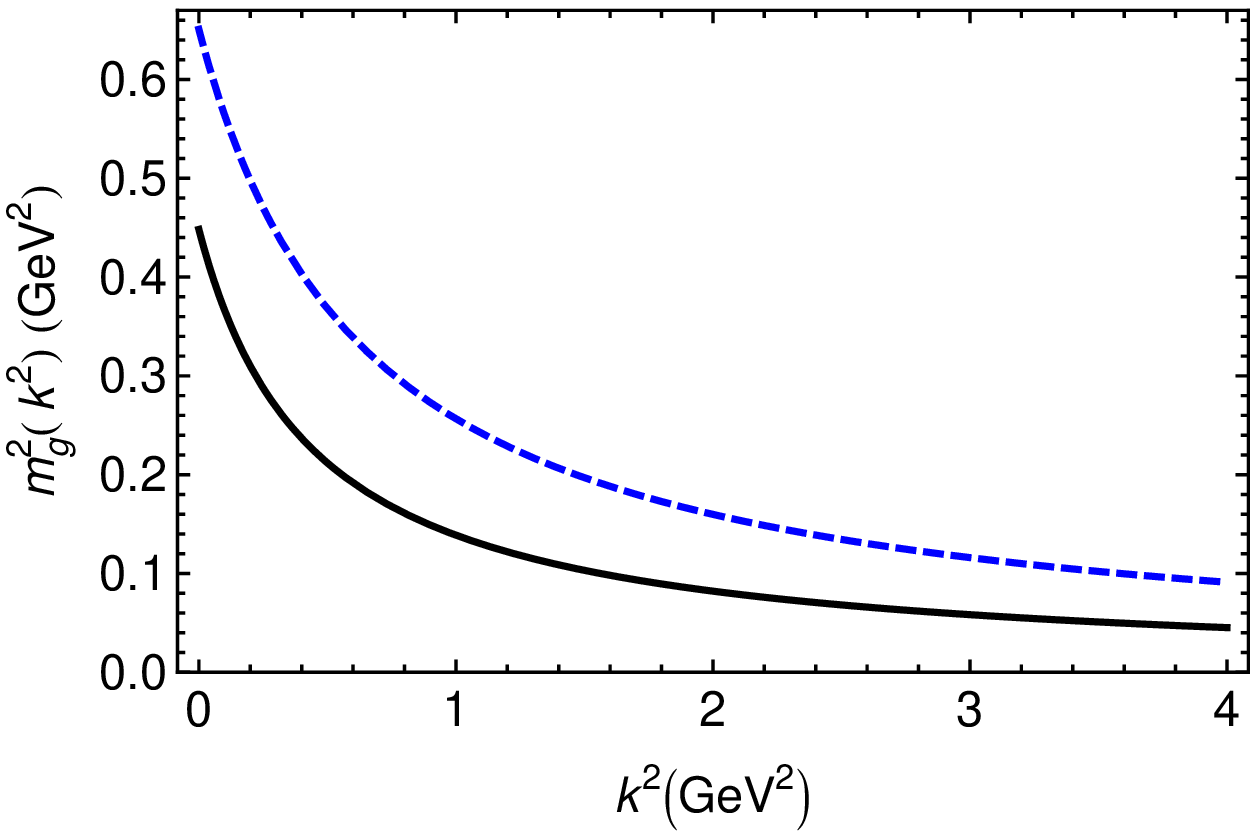}
\includegraphics[clip,width=0.46\textwidth]{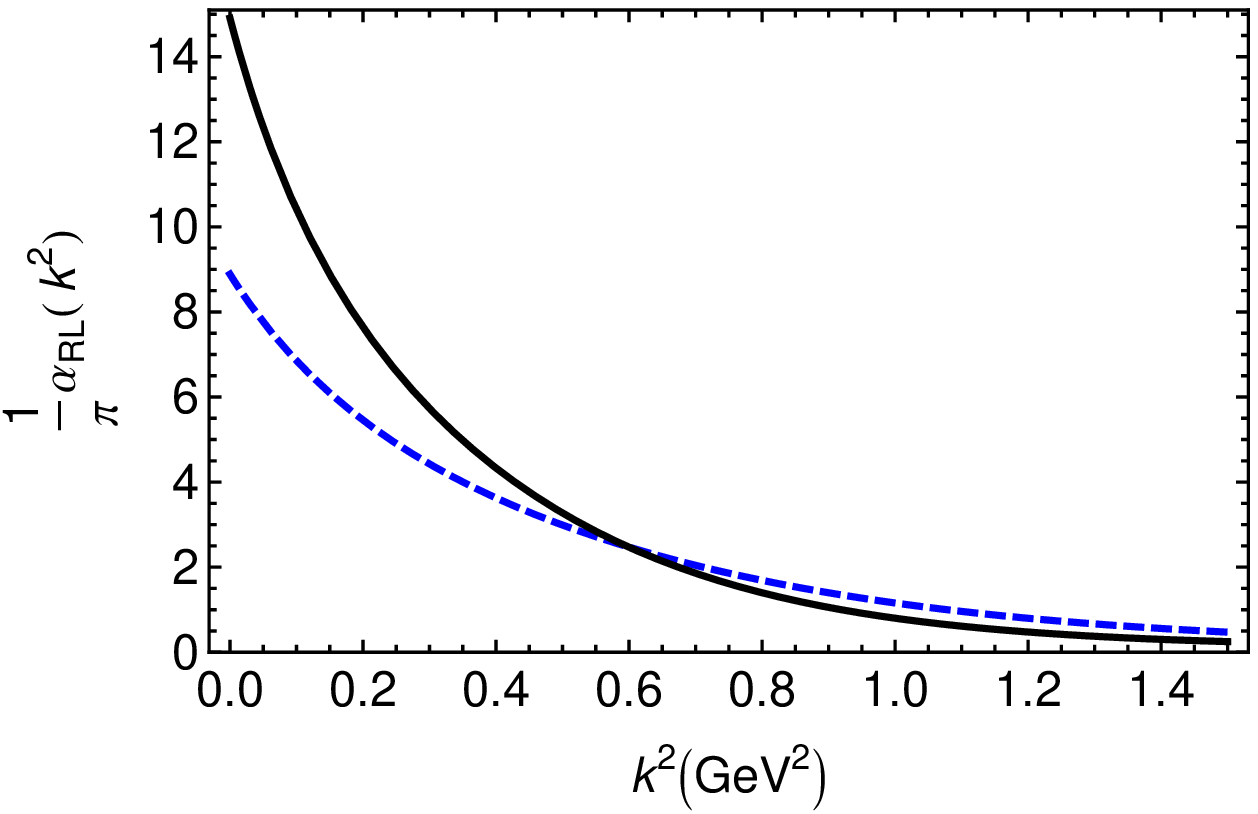}

\caption{\label{alphaeff} \emph{Upper panel} -- Rainbow-ladder gluon running-mass inferred from Eq.\,(\protect\ref{CalGQC}) via Eq.\,(\protect\ref{fitalpha}).
\emph{Lower panel} -- rainbow-ladder effective running-coupling inferred from Eq.\,(\protect\ref{CalGQC}).
In both panels: $\omega=0.5\,$GeV (solid curve); $\omega=0.6\,$GeV (dashed curve).
}
\end{figure}

However, there is reason to prefer Eq.\,(\ref{CalGQC}) over Eq.\,(\ref{CalGMT}).  Namely, its pointwise behaviour accords qualitatively with results of modern DSE and lattice studies, and it can readily be parametrised as follows \cite{Aguilar:2009nf,Aguilar:2010gm}
\begin{equation}
\label{fitalpha}
{\cal G}(k^2) \approx \frac{4\pi \alpha_{RL}(k^2)}{k^2 + m_g^2(k^2)}\,,\;
m_g^2(k^2) = \frac{M_g^4}{M_g^2+k^2},
\end{equation}
with the functions obtained in this way illustrated in Fig.\,\ref{alphaeff}.  As one would expect, the infrared scale for the running gluon mass increases with increasing $\omega$: $M_g=0.67\,$GeV for $\omega=0.5\,$GeV; $M_g=0.81\,$GeV for $\omega=0.6\,$GeV.  The values of $M_g$ are typical \cite{Bowman:2004jm,Aguilar:2009nf}.

Equally naturally, the infrared value of the coupling is a decreasing function of $\omega$: $\alpha_{RL}(0)/\pi = 15$, $\alpha_{RL}(M_g^2)/\pi = 3.8$ for $\omega=0.5\,$GeV; and $\alpha_{RL}(0)/\pi = 9$, $\alpha_{RL}(M_g^2)/\pi = 2.2$ for $\omega=0.6\,$GeV.
A context for the infrared value of the running coupling required to describe meson observables in rainbow-ladder truncation is readily provided.
With nonperturbatively-massless gauge bosons, the coupling below which DCSB breaking is impossible via the gap equations in QED and QCD is $\alpha_c/\pi \approx 1/3$ \cite{Bloch:2002eq,Roberts:1989mj,Bashir:1994az}.
In a symmetry-preserving regularisation of a vector-vector contact-interaction used in rainbow-ladder truncation, $\alpha_c/\pi \approx 0.4$ and a description of hadron phenomena requires $\alpha/\pi \approx 1$ \cite{Roberts:2011wy}.
With nonperturbatively massive gluons and quarks, whose masses and couplings run, the infrared strength required to describe hadron phenomena in rainbow-ladder truncation is unsurprisingly a little larger.
Moreover, whilst a direct comparison between $\alpha_{RL}$ and a coupling, $\alpha_{QLat}$, inferred from quenched-lattice results is not sensible, it is nonetheless interesting that $\alpha_{QLat}(0)\lesssim\alpha_{RL}(0)$ \cite{Aguilar:2010gm}.
It is thus noteworthy that if one employs a more sophisticated, nonperturbative DSE truncation \cite{Chang:2009zb,Chang:2011ei}, some of the infrared strength in the gap equation's kernel is shifted from ${\cal G}$ into the dressed-quark-gluon vertex.  This cannot materially affect the net infrared strength required to explain observables but does reduce the amount attributed to the effective coupling.

\begin{table}[t]
\begin{center}
\begin{tabular*}
{\hsize}
{|l@{\extracolsep{0ptplus1fil}}
|l@{\extracolsep{0ptplus1fil}}
|l@{\extracolsep{0ptplus1fil}}
|l@{\extracolsep{0ptplus1fil}}
||l@{\extracolsep{0ptplus1fil}}|}\hline
$\omega$ & 0.4 & 0.5 & 0.6 & $\sigma_{20}$\\\hline
$m_\pi$ & 0.214~ & 0.155~ & 0.147~ & 0.83\;\;\\ 
$m_{0^{--}}$ & 0.814 & 0.940 & 1.053 & 0.03\\ 
$m_{\pi_1} $ & 1.119 & 1.283 & 1.411 & 0.02\\ 
$m_\sigma$ & 0.970 & 0.923 & 0.913 & 1.25 \\ 
$m_{0^{+-}}$ & 1.186 & 1.252 & 1.323 & 0.34\\ 
$m_{\sigma_1}$ & 1.358 & 1.489 & 1.575 & 0.14 \\ 
$m_\rho$ & 1.088 & 1.046 & 1.029 & 1.22\\ 
$m_{1^{-+}}$ & 1.234 & 1.277 & 1.318 & 0.60\\ 
$m_{\rho_1}$ & 1.253 & 1.260 & 1.303 & 0.03\\\hline
\end{tabular*}
\end{center}
\caption{\label{tableradial}
Masses obtained with Eq.\,(\protect\ref{CalGQC}) and $D\omega = (1.1\,{\rm GeV})^3$.  The subscript ``1'' indicates first radial excitation.  The last column measures sensitivity to variations in $r_c:=1/\omega$:  $\sigma_{20}\ll 1$ indicates strong sensitivity; and $\sigma_{20} \approx 1$, immaterial sensitivity. 
Dimensioned quantities are reported in GeV.}
\end{table}

We also used Eq.\,(\ref{CalGQC}) to compute the masses of selected $J=0,1$ radial excitations and exotics, with the results presented in Table~\ref{tableradial}.  The last column in the Table was prepared as follows.  We fit the entries in each row to both $m(\omega)=\,$constant and
\begin{equation}
\label{eq:momega}
m(\omega) = \omega (c_0 + c_1 \omega),
\end{equation}
then compute the standard-deviation of the relative error in each fit, $\sigma_0$ for the constant and $\sigma_2$ for Eq.\,(\ref{eq:momega}), and finally form the ratio: $\sigma_{20}=\sigma_2/\sigma_0$.  In preparing the table we used $D\omega = (1.1\,{\rm GeV})^3$.  This has the effect of inflating the $\pi$- and $\rho$-meson ground-state masses to a point wherefrom corrections to rainbow-ladder truncation can plausibly return them to the observed values \cite{Eichmann:2008ae}.  In this connection it is notable that the value reported for $m_\sigma$ in Table~\ref{tableradial} matches estimates for the mass of the dressed-quark-core component of the $\sigma$-meson obtained using unitarised chiral perturbation theory \cite{Pelaez:2006nj}.

We have seen that with Eq.\,(\ref{CalGQC}) ground-state masses of light-quark pseudoscalar- and vector-mesons are quite insensitive to $\omega\in[0.4,0.6]\,$GeV.  Any minor variation is described by a decreasing function.  We emphasise that, for reasons which are understood \cite{Chang:2009zb,Chang:2011ei},  the behaviour of scalar- and axial-vector mesons is not well described in the rainbow-ladder truncation and hence the results reported herein do not allow reliable conclusions to be drawn about the $\omega$-dependence of their masses.

In the case of exotics and radial excitations, the variation with $\omega$ is described by an increasing function and the variation is usually significant.  (That is also the case with Eq.\,(\ref{CalGMT}) \cite{Krassnigg:2009gd,Holl:2004fr,Krassnigg:2009zh}.)  This is readily understood.  The quantity $r_\omega:=1/\omega$ is a length-scale that measures the range over which ${\cal G}_{\rm IR}$ acts.  For $\omega=0$; i.e., the $\delta^4$-function case, this range is infinite, but it decreases with increasing $\omega$.  One expects exotic- and excited-states to be more sensitive to long-range features of the interaction than ground-states and, additionally, that their masses should increase if the magnitude and range of the strong piece of the interaction is reduced because there is less binding energy. (Recall $\alpha_{RL}(0)$ is a decreasing function of $\omega$.)

Table~\ref{tableradial} confirms a known flaw with the rainbow-ladder truncation; viz., whilst it binds in exotic channels, it produces masses that are too light, just as it does for axial-vector mesons.  Plainly, no predictions for exotics can be considered reliable unless the same formulation produces realistic predictions for axial-vector masses.
It is similarly noticeable that $m_{\pi_1}$ is far more sensitive to variations in $\omega$ than is $m_{\rho_1}$; and although $m_{\pi_1}<m_{\rho_1}$ for $\omega=0.4\,$GeV, the ordering is rapidly reversed.  Thus, in conflict with experiment, one usually finds $m_{\pi_1}>m_{\rho_1}$ in rainbow-ladder truncation.
This, too, is a property of the truncation, which is insensitive to the details of ${\cal G}(k^2)$; e.g., the same ordering is obtained with a momentum-independent interaction \cite{Roberts:2011cf}.

It is probable these failings can be explained by the action of material, essentially nonperturbative corrections to the rainbow-ladder truncation which amplify effects that in quantum-mechanics would be described as spin-orbit interactions \cite{Chang:2009zb}.
For example, the $\rho_n$-mesons possess nonzero magnetic- and quadrupole-moments.  This suggests that there is appreciably more dressed-quark orbital angular momentum within these states than within $\pi_n$-mesons.  Spin-orbit repulsion could significantly boost $m_{\rho_1}$ and thereby produce the correct level ordering.
In addition, the quark model describes axial-vector mesons as $P$-wave states and it has already been established that the nonperturbative kernel corrections greatly boost their masses \cite{Chang:2011ei}.
Finally, exotic states appear as poles in vertices generated by interpolating fields with odd ``time-parity'' \cite{Burden:2002ps}, a feature which magnifies the importance of orbital angular momentum within these states.

We explored the efficacy of a form for the rainbow-ladder kernel at spacelike momenta which is a bounded, regular function that achieves its maximum value at $k^2=0$; viz., Eqs.\,(\ref{rainbowdse}), (\ref{CalGQC}).  In every respect tested, this form produces results for hadron observables that are at least equal to the best otherwise obtained in a comparable approach.  Given that the form we proposed is consonant with contemporary DSE- and lattice-QCD results on the nature of the gap equation's kernel, in future studies it should replace other \emph{Ans\"atze} that fail in this respect.
Justified, too, is repetition of numerous extant calculations; in particular, those relating to
features of the phase transition at nonzero temperature and chemical potential, such as the novel effects discussed in Refs.\,\cite{Qin:2010nq,Qin:2010pc},
and properties of excited states other than their masses.  In the latter connection, there is certainly much to add.  Herein, however, we will only report that, with $D\omega=(1.1\,{\rm GeV})^3$ and our favoured value of $\omega=0.6\,$GeV,
\begin{equation}
f_{\rho_1} = -0.45\,f_\rho.
\end{equation}
The sign is correct and significant, as argued in Ref.\,\cite{Holl:2004fr}.

More important than repeating calculations, however, is movement beyond the rainbow-ladder truncation and its incremental improvement.  Our results emphasise again that, within this circumscribed horizon, the outstanding questions in hadron spectroscopy, for example, do not have an answer.  In order to use extant and forthcoming data as a tool with which to constrain the nature of QCD's $\beta$-function, it is necessary to employ kernels that can more realistically handle systems other than ground-state vector- and flavoured-pseudoscalar-mesons.

We acknowledge valuable input from J.~Rodriguez-Quintero and S.\,M.~Schmidt.
This work was supported by:
National Natural Science Foundation of China, under contract nos.~10705002 and 10935001;
U.\,S.\ Department of Energy, Office of Nuclear Physics, contract no.~DE-AC02-06CH11357;
and Forschungszentrum J\"ulich GmbH.

\end{document}